\documentclass[11pt]{article}
\usepackage{graphicx}

\input{epsf}

\title{\Large\bf Parallel Generation of Massive Scale-Free Graphs}
\author{Andy Yoo \and Keith Henderson}
\date{Lawrence Livermore National Laboratory \\ Livermore, CA 94551}
\begin{document}
\maketitle
\begin{abstract}
One of the biggest huddles faced by researchers studying algorithms for massive graphs 
is the lack of large input graphs that are essential 
for the development and test of the graph algorithms.
This paper proposes two efficient and highly scalable parallel graph generation algorithms
that can produce massive realistic graphs to address this issue.
The algorithms, designed to achieve high degree of parallelism by minimizing inter-processor
communications, are two of the fastest graph generators which are capable of generating scale-free graphs
with billions of vertices and edges.
The synthetic graphs generated by the proposed methods possess the most common
properties of real complex networks such as power-law degree distribution, small-worldness,
and communities-within-communities.
Scalability was tested on a large cluster at Lawrence Livermore National Laboratory.
In the experiment, we were able to generate a graph with 1 billion vertices and 5 billion edges in less than 13
seconds.  To the best of our knowledge, this is the largest synthetic scale-free graph reported in the literature.
\end{abstract}
\section{Introduction}
\label{sec:intro}

Recent studies have revealed that many real-world graphs belong to a special class
of graphs called {\em complex networks} (or graphs). Examples of the real-world complex
graphs include World-Wide Web~\cite{broder:2000}, 
Internet~\cite{faloutsos99powerlaw,govindan00:heuristics,satorras2001:dynamical,vazquez2002:large}, 
electric power grids~\cite{watts98:collective},
citation networks~\cite{krapivsky2000:connectivity,redner1998:how,tsallis2000:are,vazquez2001:statistics}, 
telephone call graphs~\cite{aiello2000:arandom}, and e-mail network~\cite{ebel2002:scale-free}.
These graphs typically carry a wealth of valuable information for their respective domains.
Therefore, a great deal of research effort has been concentrated on developing algorithms
to identify and mine certain knowledge or data of interest from these graphs.
For example, algorithms that can find groups of vertices that have strong associations 
between them (called communities) have been 
reported~\cite{newman:2004d,duch:2005,newman:2004b,newman:2004c,newman:2004a}.
There exist algorithms which, given a template pattern, can find subgraphs that closely match to
the input pattern~\cite{kuramochi2001:frequent}.  Such algorithms can play a very important role in
detecting certain criminal activities or making critical business decisions.

The real-world complex graphs are typically very large (with millions or more vertices) and their sizes grow
over time.  Some researchers predict that the size of these graphs will eventually reach
$10^{15}$ vertices~\cite{korda2004:datasciences}.
The high complexity of the graph algorithms, combined with the 
large and increasing size of the target graphs, however,
makes these applications to be very difficult to apply to large real graphs.
Efforts are being made to parallelize these applications~\cite{yoo2005:ascalable,bader2006:designing}
and develop efficient out-of-core graph algorithms~\cite{sibeyn2002:heuristics}
to cope with the technical challenges.

One of key issues in developing these graph applications is the availability of
large input graphs, as these graphs are essential for the developers to
develop and test the applications and to measure their scalability and performance.
Unfortunately, we do not have publicly available real graphs
that are large enough to test the functionality and true scalability of the graph applications.
%\footnote{We consider a graph with more than one billion vertices to be large in
%this paper. Though arbitrary, the selection of this size as a yardstick can be justified by the fact that
%graphs with billions of vertices are rare.} 
A social network graph derived from the
World Wide Web, for example, contains 15 million vertices~\cite{faloutsos:2004} and
the largest citation network available has  two million vertices~\cite{newman:2001b}.
Although these real-world graphs tend to grow in size, it is unlikely that 
the real graphs of sufficiently large size will be available in the near future.

The lack of the large graphs has forced the researchers to use synthetically generated
random graphs, which are relatively cheap to construct, in their experiments~\cite{yoo2005:ascalable}.
The random graphs (also known as Erd{\"o}s-R{\'e}nyi random graphs~\cite{erdos1959:on}), however, 
is uninformative, since the structure of the random graphs 
greatly differs from that of real-world graphs.
In the absence of the large real graphs, synthetic graphs may be used for 
the development of the graph applications.
There exist several good models to synthetically
generate complex networks~\cite{barabasi1999:emergence,leskovec2005:mathematically,watts98:collective}. 
%The synthetic graphs generated by most of the models, however, fail to capture all the characteristics
%of real graphs. 
A serious drawback of these models is that they are all sequential 
models and hence, are inadequate to use to generate the massive graphs with billions of vertices and edges.

In this paper, we propose two efficient and highly scalable graph generation methods. 
Based on serial models~\cite{barabasi1999:emergence,leskovec2005:mathematically}, these methods
are designed to generate massive scale-free graphs {\em in parallel} on distributed parallel
computers.
These parallel generators require very little inter-processor communications
and thus achieve high degree of parallelism.
The first method, called {\em parallel Barabasi-Albert} (PBA) method, iteratively builds a graph using
a technique called {\em two-phase preferential attachment}.
The second parallel method, called {\em parallel Kronecker} (PK) method, applies the concept of 
{\em Kronecker product of matrices}~\cite{horn1991} and 
constructs a graph recursively in a fractal fashion from a given seed graph.
%Basically, these methods differ in that PK and PBA methods take top-down and bottom-up approaches, respectively 
These are two of the fastest graph generation algorithms with capability of generating scale-free graphs
with billions of vertices and edges. 
We have demonstrated their scalability by 
constructing massive graphs on a large cluster at Lawrence Livermore National Laboratory.
In the experiment, we have generated a scale-free graph with 1 billion vertices and 5 billion edges in less than 13 seconds,
and to the best of our knowledge, this is the largest synthetic scale-free graph ever reported in the
literature.
We also have analyzed the properties of the graphs generated by the proposed methods and report the
results in this paper.
We have found that these graphs possess commonly known properties of
real-world complex networks, including power-law degree distribution, small-worldness, and communities-within-communities.

%and evaluate two highly efficient parallel graph generation methods
%that can generate massive synthetic scale-free graphs with tens and hundres of billions of vertices, 
%in the order of seconds.
% depending on the performance .
%while preserving all the characteristics of the scale-free graphs.
%The parallel models, based on existing graph generation models~\cite{Barabasi,Falutsos},
%are developed for distributed parallel computers like clusters.
%They are very efficient in that they require very little inter-processor communications and 
%minimizes the memory accesses. Furthermore, the graphs generated by these models
%possess all the characteristics of real scale-free graphs such as power-law degree distribution, 
%small diameter, and community-within-community graph structure {\bf (what are the wish list?)}.

The remaining of the paper is organized as follows. Section \ref{sec:related-work} surveys the related work in the literature.
The proposed parallel models are described in Section \ref{sec:parallel-models}. Section \ref{sec:results}
presents the results from performance and characterization study, followed by
concluding remarks and directions for future work in Section \ref{sec:conclusions}.
\section{Related Work}
\label{sec:related-work}

Erd{\"o}s and R{\'e}nyi have proposed a simple model that generates equilibrium random graphs, called
Erd{\"o}s-R{\'e}nyi random graphs~\cite{erdos1959:on}.
In this model, given a fixed number of vertices, a graph is constructed by connecting 
randomly chosen vertices with an edge repeatedly until the predetermined number of edges are
obtained. This model is restrictive in that it produces only Poisson degree distributions.

Dorogovtsev et al. proposed a model that can generate graphs with fat-tailed degree 
distributions~\cite{dorogovtsev2003:principles}.
Given a random graph, this model restructures the given  graph by 
rewiring a randomly chosen end of a randomly chosen edge to a preferentially chosen vertex and 
also moving a randomly chosen edge to a position between two preferentially chosen vertices
at each step of the evolution.

The model proposed by Watts and Strogatz~\cite{watts98:collective} generates random structures
with small diameter, which has been named as {\em small-world} graphs. This model transforms a
regular one-dimensional lattice (with vertex degree of four or higher) by rewiring each edge, 
with certain probability, to a randomly chosen vertex. It has been found that, even with the small
rewiring probability, the average shortest-path length of the resulting graphs is of the order of
that of random graphs, and generate graphs with fat-tailed degree distributions.

The majority of recent models uses a method called {\em preferential attachment}~\cite{dorogovtsev2004}.
In a representative model among these, proposed by Barabasi and Albert~\cite{barabasi1999:emergence}, a new vertex 
joins the graph at each time step and gets connected to an existing vertex with probability
proportional to the vertex degree. With preferential attachment, these models can 
emulate the dynamic growth of real graphs.

Leskovec et al.~\cite{leskovec2005:mathematically} have proposed a graph generation model that addresses
some of recently discovered properties of time-evolving graphs: {\em densification} and 
{\em shrinking diameter}. The main idea of their model is to recursively create self-similar graphs
with certain degree of randomness. The self-similarity of the graphs is achieved by using
the {\em Kronecker product} (also known as tensor product)~\cite{horn1991}, which is a natural
tool to construct self-similar structures.
Given a seed graph, at each step this model computes the
Kronecker product of two matrices that represent the seed graph and the graph generated in previous
step respectively. 
The graphs generated with this method have regular structure. The model changes the entries in
the target matrix with a certain probability before each multiplication to 
add randomness to  the graph.

%1 H. Eves, Elementary Matrix Theory, Dover publications, 1980.
%2 T. Kailath, A.H. Sayed, B. Hassibi, Linear estimation, Prentice Hall, 2000

\section{Proposed Graph Generation Methods}
\label{sec:parallel-models}

\subsection*{Parallel Barabasi-Albert (PBA) method}
\label{subsec:para-barabasi}

Scale-free graphs can be easily generated using a well-known technique called 
preferential attachment~\cite{dorogovtsev2004}.
In a simple serial model known as Barabasi-Albert (BA) model~\cite{barabasi1999:emergence}, 
a scale-free graph is constructed, starting with a small clique, by repeatedly creating a
vertex and attach it to one of the existing vertices with probability proportional
to its current degree.

We have parallelized the BA model in this research and propose a graph generation
algorithm called {\em parallel BA} (PBA) method.
In this method, vertices are distributed to the processors,
%\footnote{Terms processor and process are used interchangeably in this paper.}, 
and all the edges adjacent to a given vertex
are stored on the same processor to which the vertex is assigned.

Sets of processors called {\em factions} are used in the PBA method.
Each processor belongs to one or more factions. The number of processors in each faction varies.
Such variation is essential for the correct implementation of
the preferential attachment operation in a distributed environment.
Furthermore, we can assign the processors to factions in a manner to enable us to
generate graphs with certain structures.
The size of each faction is a degree of freedom in this method.
The number of factions is another degree of freedom.
To facilitate the implementation, we choose to assign all vertices on a single processor to the same set of factions. 
In other words, if two vertices reside on the same processor, then they are members of the same set of factions.

It is crucial to use an efficient implementation of the preferential attachment
to allow this method to scale.
% to the graphs of massive size.
This can be done most efficiently by selecting an existing edge from the graph with a uniform probability 
and then randomly selecting one of its endpoints as the point to which a new vertex can be attached.
Therefore, an edge can be added in constant time in this implementation.

A slight variation of this algorithm is used in the PBA method.
The proposed PBA algorithm is described below in detail.
It is assumed that the algorithm runs on a processor $p$. Other processors perform
the same algorithm. We also assume that $p$ is a member of factions 
$F_0$, $F_1$, \ldots, $F_{n-1}$.
%Here, we describe the local algorithm performed by a single process ($p$).
%All the other processors execute the same algorithm in parallel.  
%
%In our implementation, given a new vertex, $v$, an edge is added to the graph by
%selecting an existing edge with a uniform probability and then connecting $v$ to one of the
%endpoints of the edge.
%instead of maintaining the current degree of each vertex, 
%Given a new vertex, its adjacent 
%We will employ preferential attachment twice in the PBA method, so it is important that we choose an 
%efficient implementation. In particular, to build a scale-free graph serially, with no community structure, 
%start with a small clique. 
%Add vertices one at a time, and for each vertex add some number of edges, k, to the graph. 
%To choose the terminal vertex for an edge, select a vertex from the graph with probability proportional 
%to its current degree. 
%Stage One: selecting processes

In the PBA method, an edge is attached in two phases.
In the first phase of our preferential attachment, 
$k$ edges are added per newly created local vertex (a vertex that resides on $p$) as in the conventional BA model.
However, each edge, $e$, associates a local vertex with
some processor $q$, instead of connecting two vertices as in the serial model.
The particular vertex that is to be the eventual endpoint of $e$ is determined remotely by the processor $q$.
%$q$\footnote{Note that $p$ and $q$ can be the same processor.}.

The processor $q$ is selected using a variation of the preferential attachment algorithm as follows.
Let $A$  denote a local edge list maintained by the processor $p$.
First, we initialize $A$ by associating the first $s$ edges with 
the processors in factions $F_0$, $F_1$, \ldots, $F_{n-1}$, 
matching sequentially one edge to one processor in the set of factions.
Here, $s$ is the total number of processors
in factions $F_0$, $F_1$, \ldots, $F_{n-1}$ (i.e., $s = \sum_{i=0}^{n-1} |F_i|$).
For an edge $e_j$, where $j \geq s$, we select an existing edge from $A$
with a uniform probability (thus realizing preferential attachment) and then 
assign its associated processor to $e_j$.
This process is repeated until the predetermined number of local vertices and edges
are created on $p$.
At the end of the first phase, $p$ sends a message to each processor $q$
to notify the number of occurrences of $q$ in $A$.

%An edge $e_j$ terminates with process $A[j]$.
%Once all the elements of $A$ have been assigned to the first $s$ edges, where $s$ denotes the number of 
%elements in $A$,
%use the preferential attachment method. That is, to 
%pick the process associated with edge s, simply pick one of the edges (0...s-1) with uniform probability. 
%The process associated with that edge will be assigned to edge s. For edge s+1, 
%choose an edge from (0...s) and assign the process similarly. 
%Repeat the process until n*k edges have been generated.

%Stage Two: selecting local vertices
In the second phase,  $p$  determines the endpoints
for the edges on remote processors and 
connects the endpoints calculated by remote processors to its local vertices.
The processor $p$ first receives messages from other processors, 
which contain the numbers of occurrences of $p$ in their respective local edge list. 
That is, the message received from a processor $q$ represents the number of 
incomplete edges one of whose endpoints resides on the processor $q$. These edges are to be 
connected to the local vertices on $p$, selected by using the standard preferential attachment technique.
Once the list of the vertices for the attachment is determined, it 
is divided up among the processors. Here, each processor is assigned as many vertices as it requested.
The selected vertices are then sent to the corresponding processors.

Having sent the endpoints for the remote edges, then $p$ receives the lists of endpoints 
from other processors for its own incomplete edges. Using the remote
vertices received, $p$ completes its local partition of the graph.
This is done by simply substituting each occurrence of processor $q$ in $A$ with the next endpoint in the list sent by $q$.
The resulting collection of edges defines the portion of the graph stored on $p$.

\begin{figure}
\centering
\begin{tabular}{c}
\begin{tabular}{|c|c|c|c|c|c|c|c|c|c|c|} \hline
$u$ & 0 & 0 & 1 & 1 & 2 & 2 & 3 & 3 & 4 & 4 \\ \hline
$v$ & $P_1$ & $P_2$ & $P_0$ & $P_1$ & $P_1$ & $P_2$ & $P_0$ & $P_1$ & $P_0$ & $P_2$ \\ \hline
\end{tabular} \\\\
\mbox{\small (a) Edge list on $P_0$ at the end of phase 1} \\\\
\begin{tabular}{|c|c|c|c|c|c|c|c|c|c|c|} \hline
$u$ & 0 & 0 & 1 & 1 & 2 & 2 & 3 & 3 & 4 & 4 \\ \hline
$v$ & 8 & $P_2$ & $P_0$ & 7 & 5 & $P_2$ & $P_0$ & 8 & $P_0$ & $P_2$ \\ \hline
\end{tabular} \\\\
\mbox{\small (b) A snapshot of edge list on $P_0$ during the phase 2} \\\\
\end{tabular}
\caption{An example of PBA graph construction on processor $P_0$. A list of edges, ($u$, $v$) is
maintained on $P_0$, where $u$ denotes a local vertex and $v$ is an endpoint determined by remote 
processors in phase 2. In this example, three factions are used, where $F_0$ = \{$P_1$, $P_2$\}, $F_1$ = \{$P_1$, $P_3$\},
$F_2$ = \{$P_0$, $P_1$\}.  Processor $P_0$ belongs to fractions $F_0$ and $F_2$.}
\label{pba:example}
\end{figure}
%example
The two-phase preferential attachment is explained using an example in Figure~\ref{pba:example}.
In this example, we generate a graph with 5 vertices per processor and 2 edges per vertex. 
It is assumed that there are three factions,  $F_0$ = \{$P_1$, $P_2$\}, $F_1$ = \{$P_1$, $P_3$\}, and
$F_2$ = \{$P_0$, $P_1$\} and processor $P_0$ belongs to fractions $F_0$ and $F_2$.
The vertices are assumed to be evenly 
distributed among the processors so that vertices 0--4 are on $P_0$, vertices 5--8 on $P_1$, and so on.

In the first phase of the algorithm, $P_0$ selects
processors and associates them with the local vertices as shown in Figure~\ref{pba:example}.a,
where the edge list on $P_0$ is depicted.  Note that the first four processors
in the list are the ones in the factions that $P_0$ belongs to, $F_0$ and $F_2$.
The rest of the processors in the list are selected using the standard
preferential attachment technique.
At the end of phase 1, $P_0$ needs four endpoints from $P_1$ (and three endpoints from each of $P_0$ and $P_2$).
These endpoints are determined by processor $P_1$ via preferential attachment and sent 
to $P_0$ in the second phase. In this example, we assume that vertices 8, 7, 5, and 8 are sent to $P_0$.
Once receiving the list, $P_0$ simply replaces the entries marked with $P_1$ with the endpoints in the list.
This is shown in Figure~\ref{pba:example}.b.

We have found that it is useful to modify the algorithm slightly to incorporate some inter-faction 
edges. In particular, during the first phase, we occasionally select a processor that is not in any of the factions
of $p$. Such processors are chosen randomly. 
The probability of creating an inter-faction edge is an another
degree of freedom in this algorithm.

\subsection*{Parallel Kronecker (PK) method}
\label{subsec:para-kronecker}

\begin{figure}
\centering
\begin{tabular}{cc}
\begin{tabular}{c}
\includegraphics[width=1in,scale=0.75]{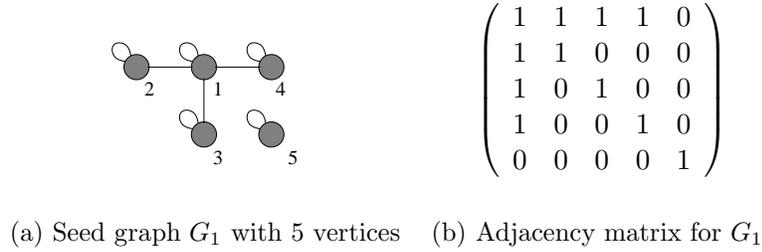} \\
%\mbox{\epsfysize=0.625in\epsfbox{seedgraph.eps}} &
%\mbox{\epsfxsize=1.25in\epsfysize=0.1in\epsfbox{seedgraph.eps}} &
\end{tabular} &
\begin{tabular}{c}
$\left( \begin{array}{ccccc} 
1 & 1 & 1 & 1 & 0 \\ 
1 & 1 & 0 & 0 & 0 \\ 
1 & 0 & 1 & 0 & 0 \\
1 & 0 & 0 & 1 & 0 \\
0 & 0 & 0 & 0 & 1 \\
\end{array}\right) $ \\\\
\end{tabular} \\
\mbox{\small (a) Seed graph $G_1$ with 5 vertices} &
\mbox{\small (b) Adjacency matrix for $G_1$ } \\\\\\
\end{tabular}

\begin{tabular}{cc}
\begin{tabular}{c}
$\left( \begin{array}{ccccc}
G_1 & G_1 & G_1 & G_1 & 0 \\ 
G_1 & G_1 & 0 & 0 & 0 \\ 
G_1 & 0 & G_1 & 0 & 0 \\
G_1 & 0 & 0 & G_1 & 0 \\
0 & 0 & 0 & 0 & G_1 \\ 
\end{array}\right) $ \\\\
\end{tabular} &

\begin{tabular}{c}
\fbox{\epsfysize=0.8in\epsfbox{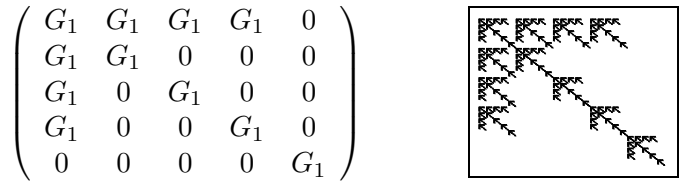}} \\\\
\end{tabular} \\\\
\mbox{\small (c) Adjacency matrix for $G_2 = G_1\otimes G_1$ } &
\mbox{\small (d) Plot for $G_4$ } \\
\end{tabular}
\caption{An example of the Kronecker multiplication for graph generation. $G_2$ is constructed
by multiplying the seed graph with itself (that is, $G_2 = G_1\otimes G_1$).
The self-similarity in $G_4$ is clearly shown in (d).}
\label{kroneckergen}
\end{figure}

A model that uses well-known concept of Kronecker matrix multiplication to generate
scale-free graphs has been recently proposed~\cite{leskovec2005:mathematically}. 
If $A$ is an $m\times n$ matrix and $B$ is a $p\times q$ matrix, then the Kronecker product 
$A\otimes B$ is the $mp\times nq$ block matrix, defined as
\begin{eqnarray*}
A\otimes B  = \left( \begin{array}{ccc}
 a_{11}B & \cdots, & a_{1n}B \\
\vdots & \ddots & \vdots \\
a_{m1}B & \cdots, & a_{mn}B 
\end{array} \right)~\cite{horn1991}.
\end{eqnarray*}
The Kronecker product of two graphs is defined as the Kronecker product of their adjacency 
matrices. 
Figure~\ref{kroneckergen} shows an example where a Kronecker
graph is generated from given seed graph using the Kronecker graph multiplication.
Since the Kronecker method is an ideal tool to construct self-similar structures, the graph
generated by this method has also self-similar structure as shown in Figure~\ref{kroneckergen}.d.

Implementing a serial algorithm for the above graph generation method is straightforward. 
Starting with an $n_0\times n_0$ adjacency 
matrix with $e_0$ edges representing a given seed graph, we recursively 
construct larger adjacency matrices using
Kronecker matrix multiplication in a top-down manner. 
To generate the $i_{th}$ matrix from the $(i-1)_{th}$ matrix, we simply replace every 1 in the $i_{th}$ matrix by 
an $n_0\times n_0$ block that is a copy of the seed graph. 
We replace every 0 
by an $n_0\times n_0$ block of zeros. So if the $i_{th}$ graph has $n_i$ vertices, 
the $(i+1)_{th}$ graph has $n_i\times n_i$ vertices. Thus, $n_i$ = $n_0^{i+1}$ for all iterations. 
We treat each edge in the graph at each iteration as a {\em meta-edge}.
A meta-edge is defined by its iteration and its position in the graph for that particular iteration. 
Given the size of a target graph, we can calculate the number of iterations  required to
generate the target graph.

A Kronecker graph can be generated efficiently by using a stack, initialized with the edges in the seed graph. 
A graph is generated by expanding meta-edges in the stack as follows.
First, a meta-edge on the top of the stack is popped up.
If its iteration, $i$, is equal to 
the predetermined final iteration, then the edge is added to the final graph. 
Otherwise, new meta-edges with iteration $i+1$ are generated and pushed onto the stack.
This operation is repeated until the stack is depleted.
We choose a stack because it guarantees that 
the memory requirement is limited to $O(\sqrt[e_0]{|E|})$, where $|E|$ is the number of edges in
the final graph. An implementation using a queue is not scalable, as it would require $O(|E|)$ 
memory space.

In the parallel implementation of the Kronecker method, 
the meta-edges are divided among the groups of processors
% called {\em processor groups}, 
at each iteration.
Each processor group generates the same meta-edges at a given iteration.
%The same meta-edges are generated at a given iteration by each {\em processor group},
%which is a subset of the processors.
If there are more processors in a processor group than there are edges in that group's portion 
of the graph, then each processor in the group is assigned to a single edge in the stack.
Here, each edge defines a new processor group that is a subset 
of the original, and  the process group ignores all other meta-edges at that iteration. 
On the other hand, if there are more meta-edges than processors in 
the processor group, the edges are divided as evenly as possible among those 
processors. Each of those processors is then in a singleton processor group for the remaining iterations.
%Stack is used in the parallel implementation as well.
Each processor must be able to calculate on the fly which meta-edges are in its 
processor group at a given iteration. 
%Meeting this requirement is straightforward and can be computed in constant time as each meta-edge is processed.

In general, it is difficult to achieve good load
balance with the PK method, as some processors may be assigned more work depending on processor group
sizes. A dynamic load balancing scheme may be used in conjunction with the PK method to overcome
this limitation.
Furthermore, some randomization logics are needed to irregulate the structure of the PK graphs.
One approach for the randomization is to add or delete meta-edges during the replacement phase at each iteration by
randomly modifying the seed graph temporarily.
% (when determining what meta-edges it generates). 
Another approach is to perform exclusive-OR operation
between the final adjacency matrix with the adjacency matrix for a random graph. 
%The former method is more structured, while the latter method introduces noise.

%Selecting the local vertices is trivial. We just assign edges $e_0$ through $e_{k-1}$ to local vertex $v_0$, 
%edges $e_k$ to $e_{2k-1}$ to vertex $v_1$, and so on. 
%To select the processes for each edge, we use a variation of the preferential attachment algorithm presented above. 
%
%Here, again we generate a scalefree graph locally. This time, use the local vertices and 
%follow the standard preferential attachment method listed above. The eventual number of edges required 
%is just the sum of all the messages received from all processes at the end of Stage One.
\section{Experimental Results}
\label{sec:results}
\subsection{Experiment environment and metrics of interest}
We have conducted a study to evaluate the proposed graph generators.
The experiments were conducted on MCR~\cite{MCRWeb}, a large
Linux cluster located at Lawrence Livermore National Laboratory.
MCR has 1,152 nodes interconnected with a Quadrics switch, and each of 
the compute nodes has two 2.4 GHz Intel Pentium 4 Xeon processors and 4 GB of memory,

In this study, we are mainly interested in evaluating the performance of
the proposed graph generators and analyzing the graphs they generate.
%The quality of a synthetic graph should be determined by
%how closely the synthetic graph matches to real scale-free graphs.
There are well-known structural and temporal properties of the 
real complex networks~\cite{leskovec2005:graphsovertime}. 
We use widely-accepted properties as indices to quantitatively evaluate the synthetic 
graph generated by the proposed methods.
%These include the power-law (heavy-tailed) degree distributions, small and shrinking diameters, and
%the property known as communities-within-communities.
%Another non-quantitative measure that must be considered here is 
%the configurability of the model to match to the size and the properties of target graph.

%degree distributions: heavy-tailed or power-law degree distributions
%small diameters
%stress
%resiliance 
%clustering coefficient
%
%shrinking diameter
%DPL

\subsection{Results}

\begin{table}
\begin{center}
\begin{tabular}{||c||c|c|c||} \hline 
Methods & $|V|$ (Million) & $|E|$ (Billion) & Time (Seconds) \\ \hline \hline
PBA	& 1,000     & 5      & 12.39		\\
PK	& 0.53    & 5.4       & 2.53		\\ \hline
\end{tabular}
\caption{Comparison of graph generation time by PBA and PK methods. The number of
vertices and edges in generated graphs are denoted by $|V|$ and $|E|$, respectively.}
\label{graph-gen-time}
\end{center}
\end{table}

Two graphs were generated by using the PBA and PK methods  on 1,000 processors on the MCR cluster,
and we report the graph generation times in Table~\ref{graph-gen-time}.
The generation time is an average of multiple runs.
We have measured the maximum time
across all processes in each run.
The disk I/O time is not included in the time reported.

Both graphs have about 5 billion edges.\footnote{We measure the size of a graph by the total space
needed to store the graph in this paper. That is, given a graph $G$ = ($V$, $E$), its size is
$|V| + |E|$. Therefore, we consider the PBA graph to be larger than the PK graph in this
experiment.}
The number of vertices in the PK graph is
considerably smaller than that in the PBA graph due to our use of a seed graph with
large average degree.
As shown in the table, it takes less than 13 seconds to generate these massive graphs.
The high generation rate can be attributed to the high degree of
parallelism of the proposed algorithms.
%The largest graph has one billion vertices and five billion edges, generated by
%the PBA method, as indicated in the table it takes about 12 seconds to generate the graph.
%Such graph generation can be attributed to the high degree of parallelism and good load-balancing 
%capability of the PBA method.
%The PK method, on the other hand, exhibits much slower performance.
%We believe that this is due to paging on some of the processors caused by load imbalance.
%That is, in the PK method some processors are assigned more edges than the others, and
%paging may take place in accessing the edges on some of the overloaded processors.
%The effect of the load imbalance is more severe when the overloaded processors are on the
%same shared-memory compute node.

\begin{figure}
\centering
\begin{tabular}{c}
\mbox{\epsfxsize=3.5in\epsfysize=3.5in\epsfbox{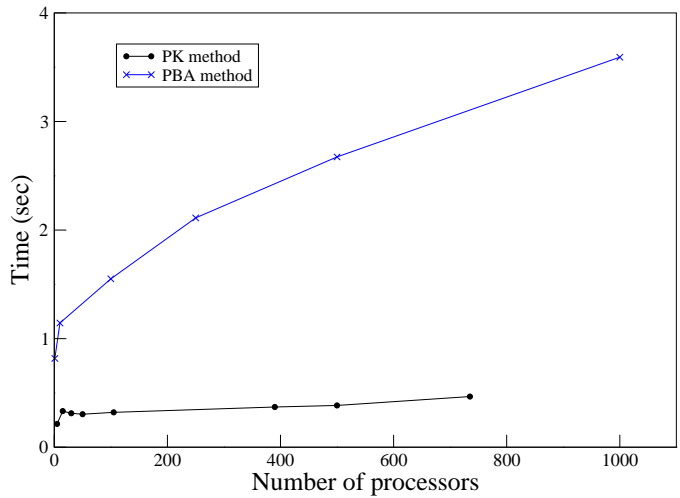}} 
\end{tabular}
\caption{Weak-scaling results}
\label{weak-scaling}
\end{figure}

The performance of both methods is further detailed in Figure~\ref{weak-scaling}, where we show
the results from a weak-scaling study.
In a weak-scaling test, the global problem size increases as the total number of processors
increases such that the size of local problem remains constant.
The local problem size of roughly one million vertices and three million edges is used.
%In generating each PK graph, different seed graph is used in order to genrate 
%a graph whose size is comparable to that of a corresponding PBA graph.
The figure reveals that the PK method is about four times faster than the PBA method.
In particular, the almost flat curve for the PK method highlights the embarrassingly-parallel
nature of the algorithm.
The graph generation time for the PBA method, on the other hand, increases as the number of
processors increases. This is because in the PBA method each processor processes endpoint vertices
sent by remote processors at the end of the execution, and the complexity of the processing
increases in proportion to the total number of processors used.
Profiling of the code confirms that each process spends most of its time in processing the received 
endpoints.
%
%In contrast to the results shown in Table~\ref{graph-gen-time}, the PK method outperforms the
%PBA method. This is because the number of edges assigned to each processor in this study
%is smaller than that in the previous experiment, and hence no paging is involved during 
%the execution. 
%Without the negative paging effect, the almost flat curve in the figure highlights the
%embarrassingly-parallel nature of the PK algorithm.
%The figure shows that the PK method is roughly four times faster than the PBA method,
%and this highlights the embarassinly-parallel nature of the PK method, where
%each processor recursively generates subgraphs without any inter-processor communications.
%The PBA method, on the other hand, requires each process to send the unattached edges it generates
%to an appropriate process that completes the attachment. 

\begin{figure}
\centering
\begin{tabular}{cc}
\mbox{\epsfxsize=3.25in \epsfysize=3.25in\epsfbox{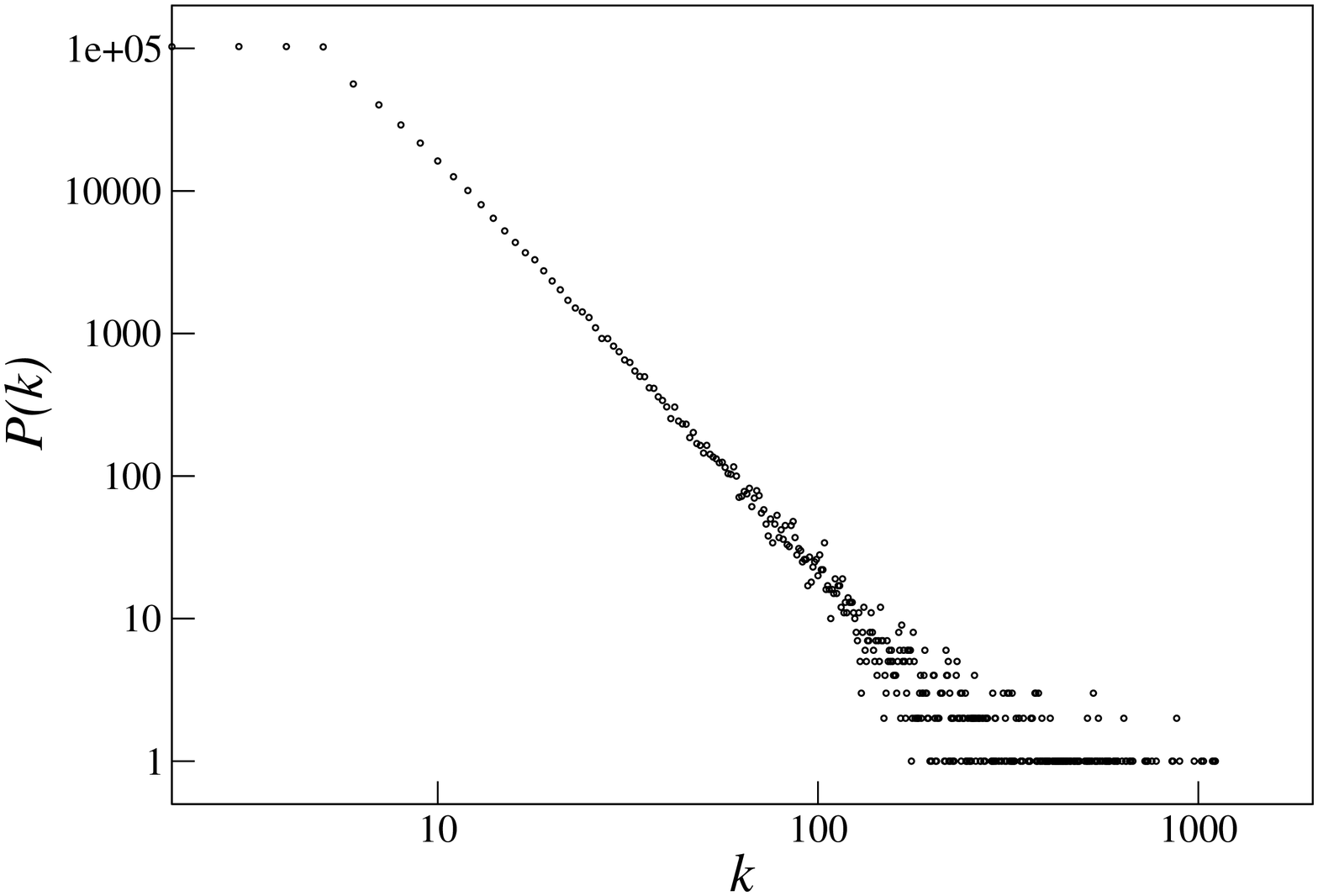}} &
\mbox{\epsfxsize=3.25in \epsfysize=3.25in\epsfbox{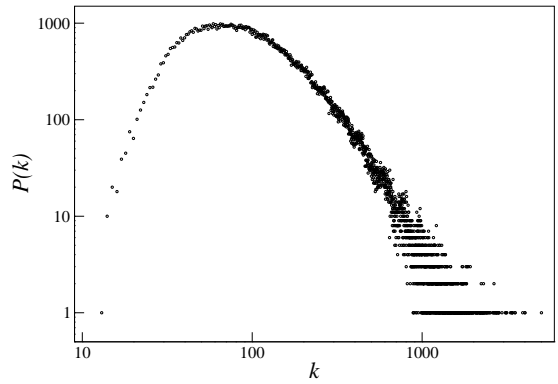}} \\
\mbox{\small (a) PBA graph ($\gamma = 2.75$)} &
\mbox{\small (b) PK graph ($\gamma = 2.47$) } \\
\end{tabular}
\begin{tabular}{c}
\mbox{\epsfxsize=3.25in\epsfysize=3.25in\epsfbox{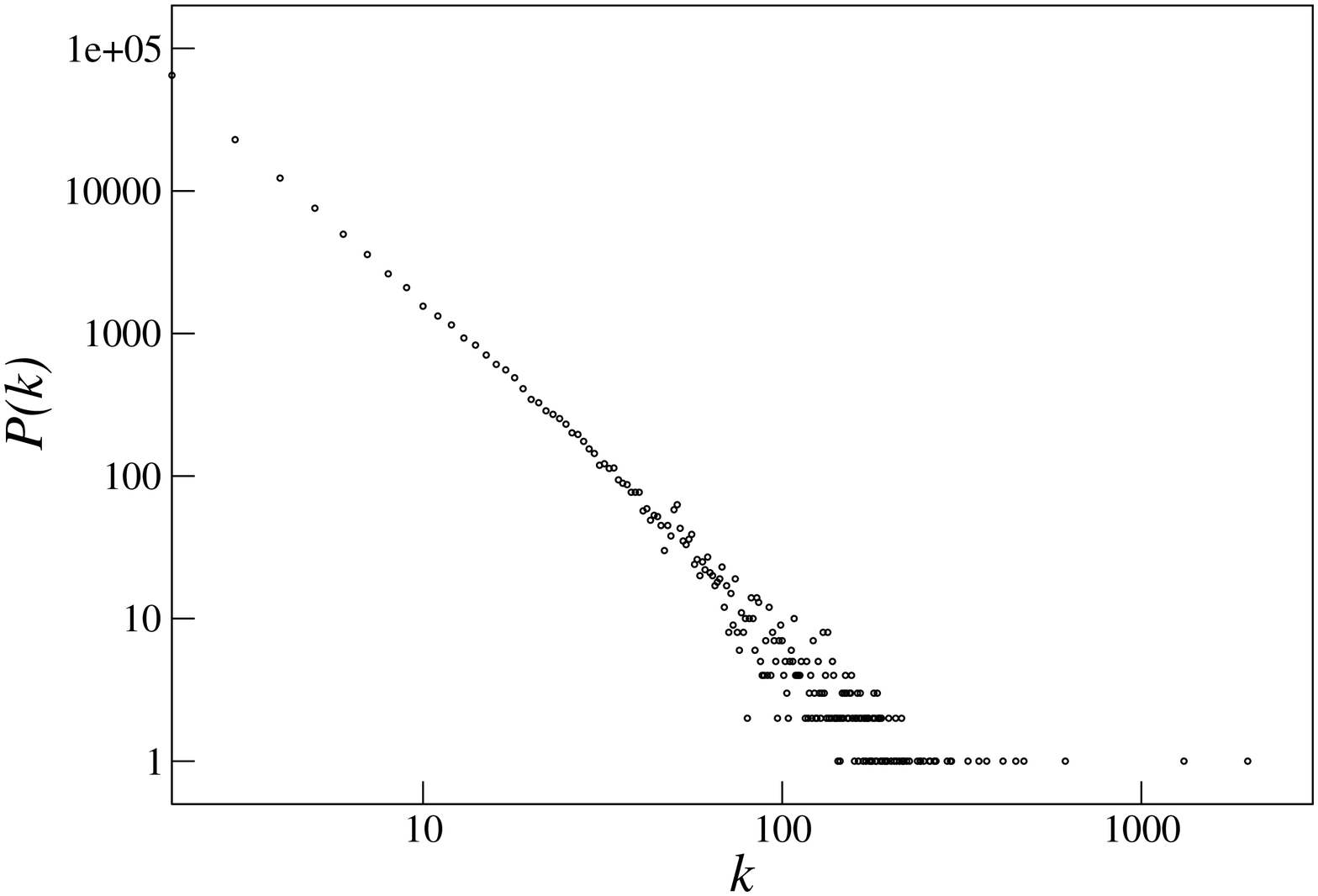}} \\
\mbox{\small (c) Router graph ($\gamma = 2.19$)} \\
\end{tabular}
\caption{Degree distributions of the synthetic and router graphs. Here $k$ and $P(k)$ denote the 
degree and the number of vertices with degree $k$, respectively.
The $\gamma$ exponent of a power-law distribution, $P(k) \propto k^{-\gamma}$, for each graph
is obtained through curve fittings and shown here as well.}
\label{degree-distributions}
\end{figure}

In the remaining of this section, we analyze the graphs produced by the proposed methods.
%A small scale-free graph with 20 vertices and the average degree of 2 is used as a graph in generating the PK graph.
A PBA graph studied in this experiments has 330,000 vertices and 2 million edges. 
A PK graph with 160,000 vertices and 28 million edges, constructed using a small seed graph with
20 vertices and 40 edges, is analyzed.
We also consider two real-world graphs, WWW and router graphs, for comparison.
%to compare some of the key properties of the real graphs to those of the synthetic graphs. 
The WWW graph has 325,000 vertices and 
2.1 million edges, and  the smaller router graph has 285,000 vertices and 861,000 edges.

Figure~\ref{degree-distributions} presents the degree distributions of the synthetic graphs and compares 
them with that of the router graph. 
The graphs are shown in a log-log scale. 
As shown in the figure, the curves  for both PBA and PK graphs are heavy-tailed. This is a signature of power-law
degree distribution that is one of the widely accepted property of real-world complex networks.
It is shown in  Figure~\ref{degree-distributions}.c that the router graph also has
fat-tailed degree distribution.
To verify that these graphs have power-law degree distributions, $P(k) \propto k^{-\gamma}$, 
we have performed curve fittings for the measured degree distributions and show the
exponent of the power-law distribution ($\gamma$) in Figure~\ref{degree-distributions}.
As shown in the figure, $\gamma$ values for the three graphs are greater than 2.
This finding coincides with the fact that if the average degree of a scale-free graph
is finite, then its $\gamma$ value should be $2 < \gamma < \infty$~\cite{dorogovtsev2004}.
The PK graph has a large number of high degree vertices.
This is because the number of low-degree vertices is small in a graph generated by the PK method,
in which the degree of a vertex grows exponentially.
We can change the degree distribution by randomly adding low-degree vertices to the final graph.
%
%The degree distribution of the PK graph, however, is not a typical of the power-law
%distribution. The discrepancy is highlighted by the small number of low-degree vertices, in particular.
%This is because with PK method in which the degree of a vertex grows exponentially
%it is hard to generate a large number of low-degree vertices.
%This limitation can be overcome by randomly adding edges at the later stage of graph generation, but
%this logic is not used when generating the PK graph.
%This is due to the fact that with the PK method the structure of the seed graph determines
%the overall structure of the final graph. 
%There are not many vertices with the degree 1 in the seed graph
%in Figure~\ref{seedgraph}.

\begin{table}
\begin{center}
\begin{tabular}{||c|c|c||} \hline \hline
Graph & Avg. Path Length & Diameter (estimated) \\  \hline \hline
WWW Graph & 7.54 & 46 \\ 
Router Graph & 8.87 & 27 \\ 
PK Graph  & 3.20 & 5 \\
PBA Graph & 6.26 & 12 \\ \hline \hline
\end{tabular}
\caption{The comparison of path length and diameter of the synthetic graphs with two
real graphs.  Both metrics are estimated by sampling to reduce the computation overhead.}
\label{patlength-diameter}
\end{center}
\end{table}

Table~\ref{patlength-diameter} presents the average path lengths and diameters of the two synthetic
graphs considered in the previous experiment as well as the WWW and router graphs. 
Both metrics are estimates obtained through sampling to reduce the computation time.
Each of the graphs analyzed has short average path length, which is the average value of the shorted path
between two randomly chosen vertices.
Further, each synthetic graph has a small diameter that is the maximum of all-pairs shortest path.
These results indicate that the graphs generated by the proposed methods have {\em small world} property,
which is another key characteristic of real-world complex networks.
Obviously, such small-worldness is more evident in the PK graph, as it contains a large number of
high-degree vertices (or hubs).
% as shown in the Figure~\ref{degree-distributions}.b.
Two real graphs, the WWW graph in particular, appear to have the  smaller number of hubs as indicated by the
larger diameters.
% in Table~\ref{patlength-diameter}.
%It should be noted that the PK graph has very average path length.  This implies the presence
%of a super-hub that is a vertex conneceted to almost all of the other vertices.
%With given seed graph,
%such a super-hub is predictable, since the structure of the seed graph determines
%the overall structure of the final graph in the PK method. 
%Natually, the PK graph should have a small diamter compared to other two graphs with the super-hub.

\begin{figure}
\centering
\begin{tabular}{cc}
\begin{tabular}{c}
\includegraphics[width=3in,scale=0.75]{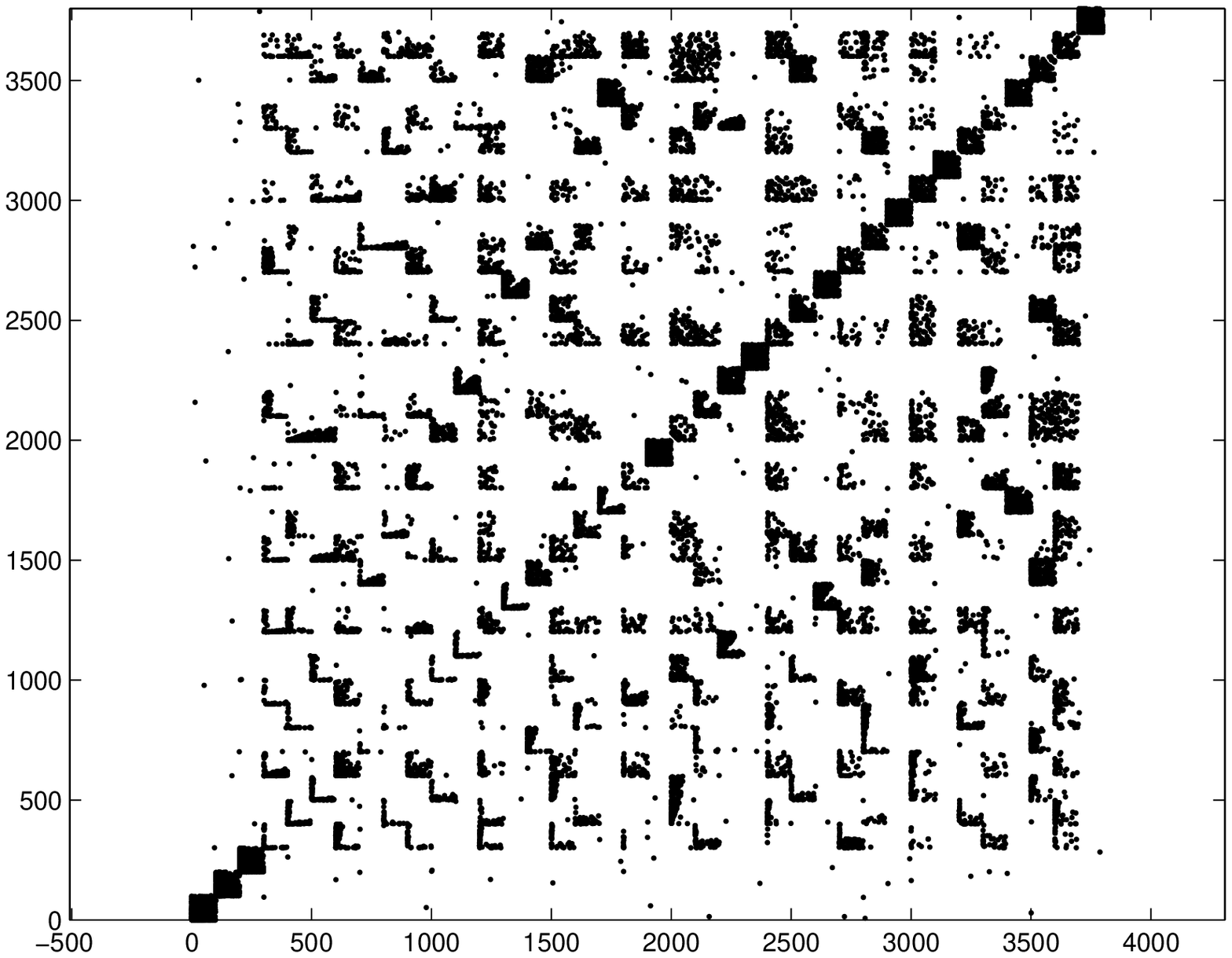}
%\mbox{\epsfxsize=3in\epsfysize=3in\epsfbox{adj-bc.eps}} 
\end{tabular} &
\begin{tabular}{c}
\includegraphics[width=3in,scale=0.75]{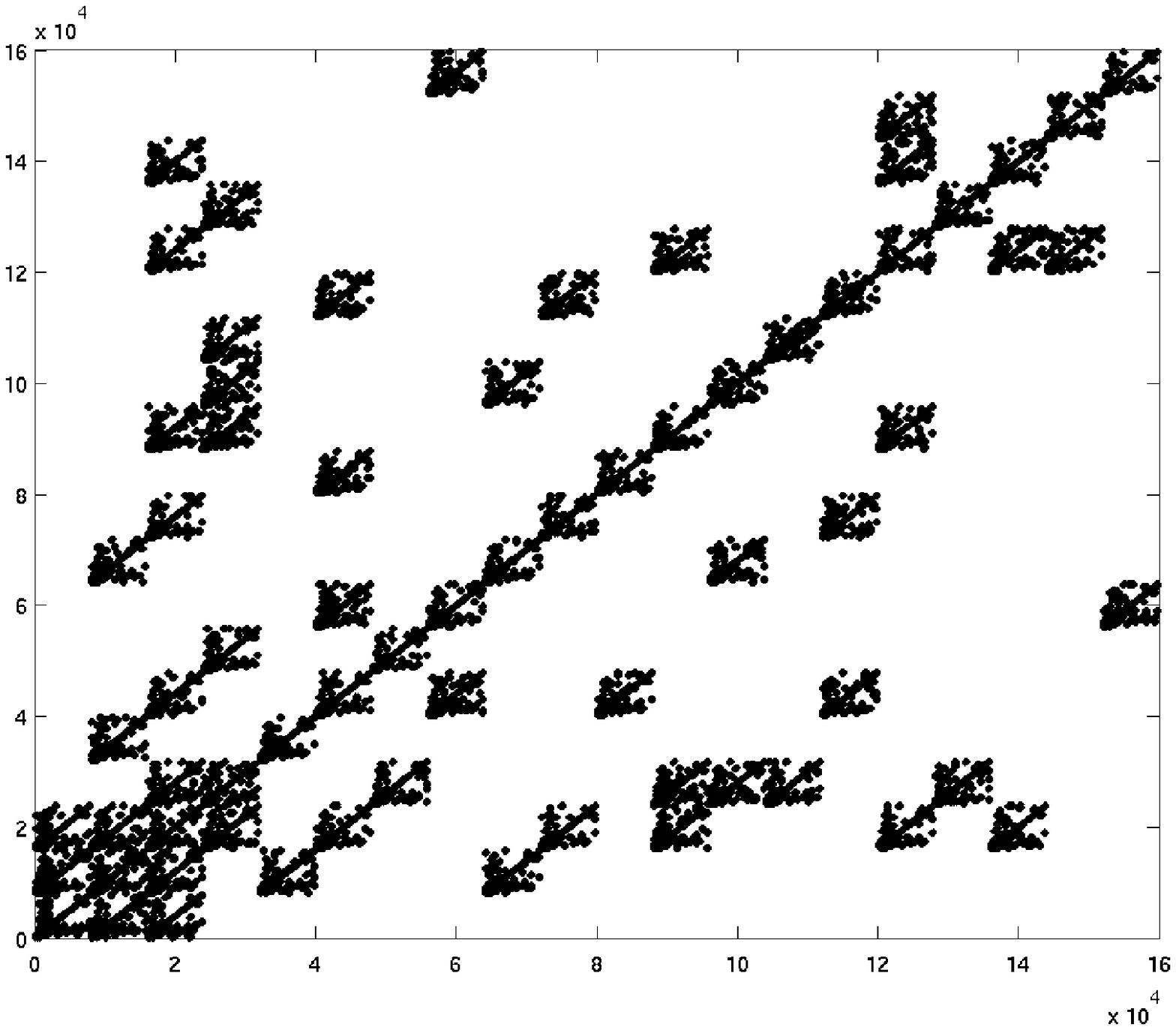}
%\mbox{\epsfxsize=2.8in\epsfysize=3.25in\epsfbox{scalefree_kron.eps}}
\end{tabular} \\
\mbox{\small (a) PBA graph } &
\mbox{\small (b) PK graph } \\
\end{tabular}
\caption{Communities within PBA and PK graphs. The graphs 
are represented as adjacency matrices.}
\label{community}
\end{figure}
In Figure~\ref{community}, we show two adjacency matrices for
PBA and PK graphs to visualize the community structures within these synthetic graphs.
As shown in the figure, the PBA and PK graphs have clearly identifiable community structures.
A major difference between the two graphs is that the PK graph has more regularly-structured
communities compared to the PBA graph.
The regular community structure of the PK graph is the result of the systematic way of
graph construction by the PK method (using the Kronecker matrix multiplication).
In addition, the self-similar nature of the Kronecker product is translated into
the communities-within-communities structure in the PK graph.
%The difference is that the PK graph has regularly structured communities, whereas
%the communities within the PBA graphs are more irregular. The regular community structure in the PK graphs
%is obviously due the systematic way of graph construction in the PK method.
%In addition, the recursive nature of the PK graph construction 
%allows communities within communities, which are clearly visible in Figure~\ref{community}.b.
%The community structure within communities is less visible for the PBA graph. 

%In reality, the real-world graphs are not as structured as the PK graph~\cite{dorogovtsev2004}, and
%rather, they resemble the graph generated by the PBA model. 
%We can randomly add and deleted some of the edges in the PK graph to irregulate its structure
%somewhat, but it turns out that such changes in the struture of the graph especially early in the
%graph construction phase results in graphs that do not have the properties of real graphs. 
%Any random addition and removal of edges at the later iterations of the graph construction will 
%have little effect on the graph structure.

\subsection{Comparison of the PBA and PK  methods}

An advantage of the PK method over the PBA method is its higher degree of parallelism.
The  PK method is embarrassingly parallel, as once a seed graph is given, 
processors generate the assigned portions of the target graph independent of each other.
A key limitation  of the PK method is that the structure of a resulting graph heavily depends on
that of the initial seed graph. 
In fact, even with randomized edge generation and removal, 
the structure of final graph largely depends on the seed graph and thus relatively regular.
In consequence, this limitation makes it very difficult to configure the PK method to generate a graph with
desired property. For example, if the seed graph is too small it is very difficult to control the degree of vertices.
To control the vertex degree, we need a relatively large seed graph, but with such a large seed graph, it is
hard to control the size of the final graph.
% since at each step the number of vertices grows exponentially to the number of vertices in the seed graph.

Although slower than the PK method, the PBA method is still a very fast algorithm.
% (as indicated in Figure~\ref{graph-gen-time})\footnote{We suspect that an incorporation of any logics into
%the PK method to overcome its aforementioned limitations will degrade its performance.}.
An obvious advantage of the PBA method is that using preferential attachment as a key means to
construct a graph, the method can be easily configured to generate  a graph of desired size and properties.
\section{Conclusions and Future Work}
\label{sec:conclusions}
Two efficient and scalable parallel graph generation methods that can generate scale-free graphs
with billions of vertices and edges are proposed in this paper.
%We have evaluated their performance.
%We have also analyzed the graphs they produce and report results.
The proposed parallel Barabasi-Albert (PBA) method iteratively builds scale-free graphs
using two-phase preferential attachment technique in a bottom-up fashion.
The parallel Kronecker (PK) method, on the other hand, constructs a graph recursively in a 
top-down fashion from a given seed graph using Kronecker matrix multiplication.
These parallel graph generators operate with high degree of parallelism.
We have generated a graph with 1 billion vertices and 5 billion edges in less than 13 seconds on a large cluster.
This is the highest rate of graph generation reported in the literature.
We have analyzed the graphs produced by our methods and shown that they have the most
common properties of the real complex networks such as power-law degree distribution,
small-worldness, and communities-within-communities.
%Also these methods serve as valuable tools for those developing algorithms for
%massive graphs, as these generators can offer large realistic input graphs which are essential 
%for developing and testing the graph algorithms.

There are other known and somewhat debatable properties of complex networks.
A rigorous study of the graphs generated by the proposed methods will reveal whether
these methods can produce synthetic graphs with these properties.
This study will also provide us with better understanding of how the logics used in our
algorithms affect the properties of the synthetic graphs they generate.
Based on this study, we will develop a set of pre- and post-generation processing and
randomization techniques that will enable us to construct a synthetic graph with
desired properties.
%
%The proposed methods currently lack 
%mechanisms to fine-tune generated graphs to mimic the properties of targeted real graphs.
%A more rigorous study of the graphs generated by these methods will reveal
%which properties of the real-world graphs are not realized by each of the proposed graph generators.
%Based on this study, a set of such fine-tuning capabilities will be added to the proposed 
%methods. In addition, 
%the interations between these fine-tuning mechanisms and their collective effect on the
%quality of generated graphs will be studied in our future work.
\bibliographystyle{abbrv}
\bibliography{references}
\end{document}